# Climate Change Attribution

# Using Empirical Decomposition of Climatic Data

Craig Loehle

National Council for Air and Stream Improvement, Inc.

552S Washington Street, Suite 224

Naperville, Illinois  60540  USA

630-579-1190

CLoehle@ncasi.org

Nicola Scafetta

Active Cavity Radiometer Irradiance Monitor (ACRIM) & Duke University

Durham, North Carolina  27708

919-660-5252

nicola.scafetta@gmail.com

**Abstract**

The climate change attribution problem is addressed using empirical decomposition. Cycles in solar motion and activity of 60 and 20 years were used to develop an empirical model of Earth temperature variations. The model was fit to the Hadley global temperature data up to 1950 (time period before anthropogenic emissions became the dominant forcing mechanism), and then extrapolated from 1951 to 2009. After subtraction of the model, the residuals showed an approximate linear upward trend after 1942. Herein we assume that the residual upward warming observed during the second half of the 20[th] century has been mostly induced by a worldwide rapid increase of anthropogenic emissions, urbanization and land use change. The warming observed before 1942 is relatively small and it is assumed to have been mostly naturally induced by a climatic recovery since the Little Ice Age of the 17[th] century and the Dalton Minimum at the beginning of the 19[th] century. The resulting full natural plus anthropogenic model fits the entire 160 year record very well. Residual analysis does not provide any evidence for a substantial cooling effect due to sulfate aerosols from 1940 to 1970. The cooling observed during that period may be due to a natural 60-year cycle, which is visible in the global temperature since 1850 and has been observed also in numerous multisecular climatic records. New solar activity proxy models are developed that suggest a mechanism for both the 60-year climate cycle and a portion of the long-term warming trend. Our results suggest that because current models underestimate the strength of natural multidecadal cycles in the temperature records, the anthropogenic contribution to climate change since 1970 should be around half of that previously claimed by the IPCC [2007]. A 21[st] Century forecast suggests that climate may warm less than 1 ºC by 2100.

## 1) Introduction

There is little doubt that the Earth has warmed since 1850, the time when global instrumental temperature estimates are first available [*Brohan et al.*, 2006]. A significant increase in the warming has been observed since 1970 relative to the period from 1940 to 1970: see Figure 1A. This sudden change in the warming trend has suggested an alarming anthropogenic effect on climate [IPCC, 2007]. However, partitioning causation has proven to be problematic. It is important to establish the relative importance of forcing factors in order to properly calibrate climate models used to project future climate scenarios and, in particular, it is necessary to determine whether multidecadal climatic patterns may be induced by natural multidecadal cycles.

Attribution studies are based on either statistical or simulation approaches, or a combination of these. Several studies [e.g., *Andronova and Schlesinger*, 2000; *Crowley*, 2000; *Damon and Jirikowic*, 1992; *Hoyt and Schatten*, 1993; *Lean et al.*, 1995; *Scafetta*, 2009; *Serreze et al.*, 2000; *Stott et al.*, 2000; *Tett et al.*, 1999] have arrived at divergent estimates of anthropogenic forcing. One reason for these divergent results is that the computations are poorly constrained. Each factor considered is uncertain both in terms of forcing and in terms of data, as documented by *Scafetta* [2009]. For example, while greenhouse gases such as $CO_2$ and $CH_4$ no doubt have a warming effect and sulfate aerosols produced by volcanoes or industrial emissions no doubt have a cooling effect, the IPCC [2007] itself acknowledges that there are large uncertainties as to the magnitude of their effects on climate. The same can be said of brown clouds (warming), black soot (warming), solar activity, and other factors. In none of these cases is a precise global dataset for the past 160 years available to provide input to an attribution study [*Kiehl*, 2007], nor is the precise forcing known independently of attribution studies [e.g., *Visser et al.*, 2000]. For example, a large uncertainty regarding the equilibrium climate sensitivity to $CO_2$ doubling is evident in Figure 9.20 IPCC [2007] WG1: the best estimate of the IPCC climate sensitivity is from 1.5 to 4.5 ºC, most likely 3 ºC, and a long tail up to about 10 ºC is also observed. Other climate mechanisms in climate models are also either poorly modeled (cloud cover, water vapor, ocean oscillations) or highly uncertain (land cover). More spatially resolved climate simulations [e.g., *Wyant et al.*, 2006] produce cloud and water vapor patterns that differ from those in operational climate models. These differences are difficult to resolve within the traditional attribution study framework.

To overcome this problem we apply an empirical decomposition method that does not depend on assumptions about forcing magnitudes, on detailed historical inputs of the various factors, nor on climate models. Our approach is to empirically characterize the pattern of global temperature change and to relate this pattern to the historical timing of greenhouse gas forcing.

## 2) Decomposition Analysis

Long-term climate trends and cycles have been detected in many geologic datasets. The most regular of these are the repeated cycles of glaciations evident in ice core data [e.g., *Rasmussen et al.*, 2006], though the precise factors governing this cycle are not fully worked out. A number of studies claim to have found decadal to centennial periodic behaviors in climate records that can be associated to solar cycles (discussed later). As well, the 11 and 22 year sunspot cycles appear to result from the effect of planetary tidal forces on the sun [e.g., Bendandi, 1931; Hung, 2007]. These findings offer the possibility that decadal to centennial scale climate could have some structure in the absence of human activity, rather than being featureless. This possibility is not ruled out by the IPCC [2007] because no scientific consensus exists on the importance of solar forcing and the current level of understanding associated with it is stated by IPCC to be low.

Using the pattern of perturbations of the sun's location from the center of mass of the solar system as a measure of the oscillations of the sun-planets system due to gravitational and magnetic interactions, which can be back-calculated by orbital calculations to any desired length of time, *Scafetta* [2010] showed via spectral analysis and other means that two particularly strong periodic signals occur with the periods of about 60 and 20 years. These oscillations are synchronous in both astronomical records and numerous global surface temperature and climate records. Spectral decomposition of the Hadley climate data showed spectra similar to the astronomical record, with a spectral coherence test being highly significant with a 96% confidence level. A model based on these astronomical cycles fit the global temperature data well and fit the ocean temperature data even better. On the contrary, the spectral patterns of climate model simulations did not match those found in the climate at all (just a 16% confidence level), suggesting that current general circulation models adopted by the IPCC do not reproduce the climate oscillations at the decadal and multidecadal scales [Scafetta, 2010].

The physical mechanisms responsible for the cycles are not understood yet and, therefore, are not included in the current climate models. However, it is reasonable to speculate that planetary tides and solar angular motion affect solar activity [*Scafetta*, 2010; *Wolff and Patrone*, 2010]. Changes of solar activity then influence the Earth's climate through changes of direct total solar irradiance, the sun's magnetic field and the solar wind. Very likely, solar activity changes influence also the intensity of cosmic rays reaching the Earth, the electrical physical properties of the terrestrial magnetosphere and of the ionosphere. The latter influences the generation of clouds which modulate terrestrial albedo and, consequently, the climate [*Kirkby*, 2007; *Tinsley*, 2008; *Svensmark et al.*, 2009]. The above empirical relationship, even if the exact mechanisms are not understood yet, offers an opportunity to detect the anthropogenic signal as a residual component not explained by natural cycles.

In what follows, we empirically estimate the structure of the temperature history data over a 160 year interval and evaluate the components of the empirical model as well as the residuals to assess natural vs. anthropogenic forcings. A key to the analysis is the assumption that anthropogenic forcings may be the predominant forcing during the second half of the 20th century with a net forcing of about 1.6 W/m$^2$ since 1950 [e.g., *Hegerl et al.*, 2001; *Thompson et al.*, 2009]. This assumption is based on Fig. 1A in *Hansen et al.* [2007] which shows that before 1970 the effective positive forcing due to a natural plus anthropogenic increase of greenhouse gases is mostly compensated by the aerosol indirect and tropospheric cooling effects. Before about 1950 (although we estimate a more precise date) the climate effect of elevated greenhouse gases was no doubt small [*IPCC*, 2007].

The approach we take is illustrated by the work of *Klyashtorin and Lyubushin* [2007] who identified a periodicity of about 60 years in long-term climate data and fit a model to the global anomaly data, with good results. A problem with their model is that it is fit to the period in which we wish to detect a difference from any natural cycles of climate. A better approach would be to fit the model to the period prior to elevated greenhouse gases and then extrapolate it forward and determine whether the forecast climate pattern corresponds to the observed one. In fact, in the latter case the model would be tested on its forecasting capability.

As explained above, we follow *Thompson et al.* [2009] and *IPCC* [2007] in identifying 1950 as the approximate year after which an anthropogenic signal appears to dominate climate forcings. *Hegerl et al.* [2001] identified 1960 as the cutoff year, but their analysis might have found an earlier year if sulfate aerosols were not included in it.

Given the strong periodicity in the solar signal detected by *Scafetta* [2010], we used this model, with two cycles of lengths 60 and 20 years, plus a linear term. A free fit of the Hadley HadCRUT3 global surface temperature data found cycles quite close to this, but we use the Scafetta solar periodicities plus a linear trend. The linear trend would approximately extrapolate a natural warming trend due to solar and volcano effects that is known to have occurred since the Little Ice Age, a period that encompassed the Maunder and Dalton solar minima of the 17-19th centuries [Eddy, 1976; Scafetta 2009].

As explained in Scafetta and West [2007] and in Scafetta [2009] most of this warming is likely due to the increased solar activity with the volcano effect playing only a minor effect if recent paleoclimatic reconstructions of the global climate, which show a large pre-industrial variability [e.g. Moberg et al., 2005], are adopted. On the contrary, only using obsolete *Hockey Stick* temperature graphs, which do not show a significant pre-industrial variability [e.g. Mann et al., 1999], it is possible to conclude that volcanic and solar effects may be equally important in explaining the Little Ice Age [Crowley, 2000; Scafetta and West, 2007].

The model was fit by nonlinear least squares estimation using *Mathematica* functions, with phase and amplitude free but period fixed as above. The validity of this approach for estimating parameters for a time series model with data of this length is demonstrated in Supplemental Information 1. When fit to the entire 160 year period, the model

$$y(t) = A \cos[2\pi (t-T_1)/60] + B \cos[2\pi (t-T_2)/20] + C(t-1900) + D \qquad (1)$$

performs only decently (Fig. 1a). However, the nonrandom residual signal over the entire period (Fig. 1b) reveals that the model is deficient. Since this mis-specification could result from not including anthropogenic effects that could have accelerated the warming since 1950, the model was refit using the data only up to 1950 (Fig. 2a). See Table 1 for the evaluated regression coefficients in the two cases.

Just as in the solar data, the 60-year cycle has about three times the peak-to-trough amplitude (~0.24ºC) of the 20 year cycle (~0.08 ºC).  The model fit is $R^2 = 0.53$.  Comparing the residuals over the entire period depicted in Fig 2b, they appear stationary up to 1950, but there is a visible positive trend from 1950 to 2010.  The residual warming observed since 1950 could be mostly induced by an increasing anthropogenic warming signal due to global industrial development, land usage change and urban heat island (UHI) effects since 1950.

A linear fit to the residuals since 1950 (Fig. 3) has a slope of 0.66 +/- 0.08ºC/century ($R^2 = 0.59$) and begins (>0) in 1942.  Note that given a roughly exponential rate of $CO_2$ increase [*Loehle*, 2010] and a logarithmic saturation effect of GHG concentration on forcing, a quasi-linear climatic effect of rising GHG could be expected.  In any case, this upward trend is the result of all positive and negative additional climatic forcings (e.g., the anthropogenic ones) and other possible contributions (e.g., poor adjustment of the temperature data) that are not implicit in our model beginning in 1942.

Figure 2a shows that the model fit from 1850 to 1950 reproduces the modulation of the temperature up to 1942 and has random residuals over this period (Fig. 2b).  It is very unlikely that the 60 and 20-year cycles would match the pattern of temperature over this period by chance; thus, the empirical model calibrated on the temperature data before 1950 has the capability of *forecasting* the multidecadal oscillations of the temperature after 1942, which we do next.

The above finding is also confirmed by two recent studies that show a similar linear anthropogenic warming trend after about 1950 [*Thompson et al.*, 2009] and about 1960 [*Hegerl et al.*, 2001].  Both studies also find a small residual warming in the pre-1950 or 1960 period but do not attribute this trend to human influences.  The magnitude of the *Thompson et al.* [2009] residual post-1950 trend (i.e., the anthropogenic component) is similar to our result.  These studies attempt to filter out sources of extraneous noise, such as ENSO events and volcanoes.  They make many assumptions, which are not needed in the present study in which short-term perturbations are simply noise relative to the 60 year cycle and the long-term natural trend.

To verify the combined effect of natural plus anthropogenic forcings, a full model was constructed, with the anthropogenic linear trend obtained by fitting the residual from 1950 to 2010 assumed to be in effect since 1942. This model (Fig. 3a) fits the entire 160-year record better than any model we have seen. Fig 3b depicts the two harmonics at 60 and 20-year period and the linear natural trend (fit 1850-1950) added to the anthropogenic linear trend (fit of the residual in 1950-2010) since 1942.

It would thus appear that the modeling approach used here has captured the essential features of climate from 1850 to 2010. A model based on 100 years of data (from 1850 to 1950) shows stationary fluctuating residuals (Fig. 2b). The extrapolation period 1950 to 2010 shows an upward linear pattern of residuals, which is likely a signal of anthropogenic warming. The slope of this anthropogenic warming trend is 0.66°C/century since 1942. Figures 3a and 3b show that about 50 +/- 10% of the 0.8°C global surface warming observed from 1850 to 2010 is likely the result of a natural warming trend recovery since the LIA plus the combined effects of 20- and 60-year natural cycles. In particular, the sudden and alarming warming increase observed from 1970 to 2000 is mostly a consequence of the combination of the 20- and 60-year natural cycles. Both cycles were at their minima in about 1970 and at their maxima around 2000, as Fig. 3b shows. These cycles would imply a combined warming of about 0.32°C on the total observed warming of about 0.5°C since 1970. Thus, about 60% of the warming observed since 1970 can be associated with natural 20- and 60-year multidecadal cycles.

**3) Multi-secular evidence of a quasi 60-year climatic cycle**

Here we list empirical studies supporting the existence of a quasi 60-year cycle in the climate system. A 20-year cycle with smaller amplitude will necessarily be harder to detect in geologic records, and is not evaluated here. Also, note that climate may be characterized by other cycles with periods longer than 60-year [e.g., *Loehle and Singer*, 2010]. Because complex interference patterns may emerge through the superposition of all cycles and volcano eruptions can further disrupt the signal, perfect 60-year harmonics may not be easily visible in the records. Noisy geologic data may not show a signal even if one is present.

While spectral coherence of the Hadley historical data and solar data has been established [*Scafetta*, 2010], does this effect extend before 1850? Before 1850 global instrumental temperature records are not available. Thus, it is possible to use only a few historical European temperature records dating back since the 17th century and several paleoclimatic records. These records may present several problems and may be a poor proxy for the real global climate.

It has been recently noted that a 500 year temperature reconstruction in the Mediterranean basin (Spain, France, Italy) by means of documentary data and instrumental observations suggests a dominant ~60 year oscillation [*Camuffo et al.*, 2010]. Moreover, there are multidecadal periods of high and low ENSO-type activity evident in historical records going back several hundred years [e.g., *Biondi et al.*, 2001; *Mantua and Hare*, 2002; *Mantua et al.*, 1997; *Minobe*, 1997, 1999, 2000; *Patterson et al.*, 2004; *Shabalova and Weber*, 1999; *Wiles et al.*, 2004; Fig. 5 in *Gergis and Fowler*, 2009]. The long-term ENSO-type signal has been shown to be quasi-periodic, with period of 50 to 70 years, in records such as the Greenland ice core, bristlecone pine ring widths, and sea sediment records of fish abundance going back thousands of years [*Klyashtorin and Lyubushin*, 2003, 2007; *Klyashtorin et al.*, 2009]. Over these long periods, cycles of 50 to 70 years were shown, using moving window spectral decomposition, to increase in strength over the last 600 to 1000 years, reaching a peak strength in the 20th century. *Klyashtorin and Lyubushin* [2007] and *Klyashtorin et al.* [2009] also showed that over the past 100+ years an average 60 year global temperature cycle is strongly coherent with the Pacific Decadal Oscillation, the zonal mode of the Northern Hemisphere Atmospheric Circulation Index, the Aleutian Low Pressure Index, flood levels in the Neva River (Russia), and precipitation in Oregon, among others. In addition, the ups and downs of major ocean fish stocks are almost completely explained by this ~60 year climate cycle [e.g., *Klyashtorin and Lyubushin*, 2007; *Mantua et al.*, 1997]. The Atlantic Multidecadal Oscillation also closely tracks this approximately 60-year cycle [Fig. 2 in *Levitus et al.*, 2009]. *Wiles et al.* [2004] found evidence for the persistent effect of the PDO on climate in Alaska over the past 1000 years. A quasi-60 year periodicity is found in secular monsoon rainfall records from India [Agnihotri and Dutta, 2003]. *Yousef* [2006] showed a good anti-correlation between ENSO events and the Wolf-Gleissberg cycle over 300 years. A clear quasi 60-year cycle is observed in the global sea level rise record since 1700 [Jevrejeva et al., 2008]. A 50 to 80 year cycle over hundreds of years has also been found in sedimentary records in the NE Pacific and their cause has been related to solar and cosmic ray activity cycles [*Patterson et al.*, 2004]. *Ogurtsov et al.* [2002] found strong evidence for the cycle, such as a 60 to 64 year period in $^{10}$Be, $^{14}$C and Wolf number over the past 1000 years, which may indicate a solar origin. The solar system oscillates with a 60-year cycle due to the Jupiter/Saturn three-synodic cycle and to a Jupiter/Saturn beat tidal cycle [*Scafetta*, 2010].

For example, Figure 4 depicts two multi-secular climate records showing multiple quasi-60 year large oscillations since 1650. Figure 4a depicts the abundance of *G. Bulloides* (an indicator of surface temperatures) found in the Cariaco Basin sediments in the Caribbean sea since 1650 [*Black et al.*, 1999]: the best sinusoidal fit gives a period of about 61.5 +/- 4 years. This record, which correlates well with periods of reduced solar output, is an indicator of the trade wind strength in the tropical Atlantic Ocean and of the North Atlantic Ocean atmosphere variability.  Figure 4b depicts the reconstructed PDO based on tree-ring chronologies from *Pinus flexilis* in the United States West Coast [*MacDonald and Case*, 2005]:  the best sinusoidal fit gives a period of about 59.5 +/- 4 years.  Thus, both records show a clear 60-year cycle.

A potential source of data for evaluating this cycle can be found in multiproxy temperature reconstructions [e.g., *Moberg et al.*, 2005].  Unfortunately, most of these reconstructions must be rejected as accurate data sources: in fact, paleoclimatic temperature reconstructions appear quite different from each other [North et al., 2006].  Many rely on tree ring data, but it has been shown [e.g., *Loehle*, 2009b] that tree ring data, as well as other proxy records, may also have unresolved issues for developing reliable climate histories.  In other cases, multiple proxies either are dated at wide intervals such as 100+ years, precluding detection of a multi-decadal signal, or suffer from large dating error, which in combined series will lead to a smearing of multi-decadal peaks [*Loehle*, 2005].

Given that this ~60 year periodic climate pattern has such a long empirical history with some evidence for links to solar activity and solar system oscillations, it cannot be attributed to human activity and cannot be simply random noise. Therefore, we are justified in factoring it out of the historical record in order to discern the human influence on climate.  We next evaluate additional components that can be potentially factored out and, then, develop a final model that is extrapolated to 2100.

**4)  Tropospheric Aerosols**

A major factor often considered crucial for proper climate modeling is the influence of tropospheric aerosols that result from industrial activity, which are often assumed to have been a major cooling influence in the 1950s through the 1970s due to their reflective properties (although *Schiermeier* [2010] argues that the aerosol extent and forcing data are shaky at best). However, because the climate appears to contain a ~60 year cycle for several centuries, much of the cooling may have been induced by this cycle, which was in its descending trend during this period. In fact, the residuals from the pre-1950 model (Fig. 2b) over the decades of the 1950s through 1970s show not a decline but a constant linear rise for the anthropogenic component, which is inconsistent with a strong cooling effect over this period. Our full model (including the linear anthropogenic effect since 1940) shows a cooling in the 1950s through 1970s without any forcing from tropospheric aerosols (Fig. 3a). Note that the amplitude of this 60-year cycle has been determined fitting the data before 1950. Moreover, the periods 1880-1910 (slight cooling) and 1910-1940 (significant warming) are statistically equivalent to the periods 1940-1970 (slight cooling) and 1970-2000 (strong warming), respectively. Thus, while tropospheric aerosols might have contributed to the cooling between 1940 and 1970, most of the observed cooling is likely associated with the 60-year cycle, which was in the declining phase during that period.

## 5) Land Use Effects on Trends

The final factor considered is the urban heat island (UHI) effect, including other land use/land cover (LULC) changes [e.g., *Klotzbach et al.*, 2009]. The UHI effect has long been known [e.g., *Klysik and Fortuniak*, 1999]. It results because urban structures have lower albedo, which raises temperatures, and retain heat, which raises nighttime minima. In addition, there is less evaporative cooling in a city. Herein, we discuss the current uncertainty relative to this phenomenon. It is not known how well the land surface temperature records are corrected for the UHI effect.

The UHI effect arises not from the mere fact that a city is warmer than a rural site [e.g., *Klysik and Fortuniak*, 1999], which is of no consequence for calculating trends, but from the effect at a given weather station location as the city grows around it, causing a gradual warming bias over time [e.g., *Böhm*, 1998; *Magee et al.*, 1999].

While many studies have documented UHI effects at the city level, few have done so at larger scales. The contribution of UHI effects at larger scales arises from the fact that a large proportion of weather stations have been established to serve cities and airports, with few in remote locations. Thus, UHI effects contaminate the instrumental temperature record. In a landmark study, *Jones et al.* [2008] evaluated the UHI effect for China and found that 39.6% of the instrumental warming from 1951 to 2004 was contributed by UHI and related effects. In another set of studies, *McKitrick and Michaels* [2004, 2007] and *McKitrick* [2010] showed that local warming (deviations from global trends) was proportional to national economic activity, a surrogate for urbanization, and that this effect is real and not caused by, for example, atmospheric circulation patterns. They estimated that about half of the land warming claimed by the IPCC to be UHI and land use related. In an analysis of trend differences between satellite and surface data, *Klotzbach et al.* [2009] suggested that from 30 to 50% (based on UAH satellite data) of the instrumental land surface warming trend from 1979 to 2008 was spurious. *Kalnay and Cai* [2003] found 0.27°C per century of the warming trend over the continental United States to be due to land use change effects. Other studies [e.g., *Christy et al.*, 2006; *Fall et al.*, 2009] support these results.

The above studies suggest that the UHI problem is still unsolved and its effect may have been poorly filtered out from the global surface record despite the attempts to do so. The warming since 1950 might have been lower than what is currently believed and the climate effects of the anthropogenic forcings may have been further overestimated also for this reason. Thus, the modeled upward warming trend attributions depicted in Figure 3a should be interpreted as an upper limit for an anthropogenic (GHG plus aerosol) contribution to climate change.

**6) 21st century forecast of climate change**

Based on the analysis in this study, it is possible to make a preliminary forecast of future trends that may be reliable for the coming few decades. We believe this is possible because the natural variability may be assumed in first approximation to be a sum of a possible upward linear trend started since the Little Ice Age, which may still continue during the next decades although it is likely part of a millennial long natural cycle which includes the Medieval Warm Period and the Little Ice Age [see *Loehle and Singer*, 2010], plus the periodic multidecadal component we have detected.

That a natural trend may not exceed the above estimated upward linear trend from 1850 to 1950 may be indirectly suggested by the fact that since 1930 the sea-level acceleration has been, on average, slightly negative [Houston and Dean, 2011], which may suggest that a millenarian natural climate cycle is turning down. Interestingly, the measured deceleration of the sea-level rise from 1930 to 2010 occurred despite the strong positive acceleration of anthropogenic GHG emissions during the same period. Moreover, the anthropogenic component of the warming (GHG and aerosol forcings and/or poorly corrected LULC and UHI effects) seems to have resulted in a linear additional warming trend from 1942 through 2010 without any significant positive acceleration. The combined (20 and 60-year cycle plus two linear trends) model depicted in Figure 5 matches the 160-year historical record exceptionally well (Fig. 5). Thus, extrapolating the model another 90 years may be useful.

The result (Fig. 5) is a continued warming with oscillations to a high in 2100 of about 0.6 °C above 2000 values. However, if part of the linear warming trend is due to UHI and LULC changes and/or erroneous data adjustments, the anthropogenic effect due to GHG and aerosol would be smaller. Moreover, the full anthropogenic effect of 0.66 ºC/century would be masked by the natural multidecadal 20 and 60-year cycles until 2040. Note that for the full model the downturn in temperature evident in the satellite observations after 2000 [*Loehle*, 2009a] is present. This downturn in the temperature since 2000 was not reproduced by any projection of the climate models adopted by the IPCC [2007] although it is predicted by the celestial empirical model of climate recently proposed by Scafetta [2010]. The above estimate of a possible warming of only 0.6 °C above 2000 should be compared against the IPCC projections that claim a monotonic anthropogenic warming with an average global value in 2100 of 3 to 3.4 °C above 2000 values for the likely scenarios.

### 7) Can Solar Activity Explain the Temperature Modulation?

If anthropogenic forcing has been overestimated by the models adopted by the IPCC, it is necessary that other natural forcings such as the solar forcing have been underestimated by the same models to balance the secular warming trend observed in the temperature record.

Herein we investigate whether solar activity can explain a quasi 60-year modulation of temperature and at least part of the upward trend by expanding the discussion presented in *Scafetta* [2009]. The task is not simple, because past total solar irradiance (TSI) variation is not known with certainty. We develop an enhanced version of the *Scafetta* [2009] model in Supplemental Information 2, which uses the TSI record as a proxy for determining a phenomenological signature of the total solar effect on climate. This model empirically evaluates the climate sensitivity to solar changes and, therefore, it automatically takes into account natural amplification mechanisms such as those due to cloud and GHG feedbacks.

Figure Sup. 2.1 in Supplementary Information 2 shows that the proxy model (green) presents a slight increase between 1850 and 1880/90, a decrease from 1880/1890 to 1910, an

increase from 1910 to 1940/50, a decrease from 1940/50 to 1970/80, and an increase from 1970/1980 to 2000, and a decrease after 2000. Thus, it shows a quasi 60-year cycle, which is clearly present in the gravitational oscillations of the solar system [*Scafetta*, 2010]. The latter 30-year pattern since 1980 is perfectly compatible with the ACRIM total solar irradiance satellite measurements that shows an increase from 1980 to 2000 and a decrease from 2000 to 2010 [Scafetta and Willson, 2009]. This is an important result because some solar scientists currently believe that ACRIM trend is instrumental and not of solar origin (Frohlich, 2006; Lockwood and Frohlich, 2008).

From the above, it appears that these TSI proxy models can approximately reproduce the quasi 60-year modulation of the temperature. In addition, they contain an upward trend that is part of a multisecular solar cycle responsible for much of the cooling from the Medieval Warm Period to the Little Ice Age and the warming from the Little Ice Age to the current warm period. To better show this we applied the empirical temperature model of *Scafetta* [2009] to the novel green TSI proxy curve and to the average of the three TSI curves represented by the black curve in the Supplemental figure to obtain values for forcing. We plotted the two solar phenomenological signatures in Figure 6 against the global surface temperature record.

Figure 6 confirms that there is good agreement, especially after 1900, between the empirical ~60 year modulation of temperature (Fig. 3a thick line) and the climate signature that would be imprinted by solar multidecadal variation. If the correspondence is not rigorous it is because the available TSI proxy models can only approximately reproduce the real multidecadal TSI variation and more research should be done on this topic. It also appears that a modest upward trend over the period since 1850 of 0.11°C/century is reproduced by the TSI reconstructions. This is a critical point because the rising temperatures from 1850 to 1950 (or 1942) could be due to the secular increase of solar activity observed since the 18[th] century [Scafetta, 2009]. Thus, the cause of both the 60-year cycle modulation plus an upward trend since 1850 is likely to be linked to the multidecadal and multisecular oscillatory nature of the solar variation. This is evident over long timescales and is apparent in cosmogenic isotopes. For example, *Ogurtsov et al.* [2002] found strong evidence for solar cycles with 60 to 64, 80 to 90, 128, 205, and 1020 year periods in [10]Be, [14]C, and Wolf number over the past 1000 years. *Mazzarella* [2007] also found a clear link between the 60-year solar cycle and climate oscillations during the last 150 years.

## 8) Discussion and conclusion

Herein we have argued that the global surface instrumental temperature record, as well as longer term records, contain a quasi 60-year natural cycle. This cycle may be associated with a low frequency modulation of the ocean oscillations such as the ENSO signal. It appears to be present in solar and astronomical records as well, although a perfect match is difficult to obtain because of the uncertainties about the current total solar irradiance proxy models. Better solar irradiance models or alternative solar system direct forcings [Scafetta, 2010] may be required to find a perfect match.

Thus, this quasi 60-year cycle observed in the temperature record likely has an astronomical/solar and, therefore, natural origin. The weaker quasi 20-year cycle is not as easily detected in geologic records but it is clearly detected in the global surface instrumental records [*Scafetta*, 2010]. The projected rate of temperature increase due to human influences in our analysis suggests a value for climate sensitivity. Using an estimate of doubling time of greenhouse gases based on historical doubling times for carbon dioxide [*Loehle*, 2010] as a proxy, the anthropogenic warming resulting from doubled forcing would be two to three times less than previously thought, that is on average about 1-1.5°C from our linear model, which may also suggest the presence of a slight negative feedback to $CO_2$ in the climate system.

The models adopted by the *IPCC* [2007] assume that more than 90% of the warming observed since 1850 to 1900 and about 100% of the warming observed since 1970 has been caused by anthropogenic forcing. This information can be deduced from a direct analysis of the figures published there, such as figures 9.5a and 9.5b, which claim that natural forcing alone (solar plus volcano) would have induced a cooling since 1970. However, these climate models are not able to reproduce decadal and multi-decadal natural cycles [Scafetta, 2010]. We showed that the Hadley global anomaly data from 1850 to 1950 are matched closely by a periodic model with periods 60 and 20 years plus a linear trend. The residuals from this model after 1950 are strictly linear, matching other attribution studies [e.g., *Hegerl et al.*, 2001; *Thompson et al.*, 2009]. Based on this result, our full model (natural plus anthropogenic effects after 1942) closely matched the full 160-year record. It had far higher pattern accuracy than any traditional analytical computer general circulation climate model output. We further found no clear evidence for a significant cooling effect due to troposphere aerosols, since our model shows cooling in the 1950s through 1970s without invoking this mechanism, and there are no unexplained residuals corresponding to the period of assumed influence of this factor. Comparing the temperature rise from the 1970s to 2010 (~0.5°C) due to the 60 and 20-year cycles (~0.3°C) with that evident in the model residuals (Fig. 2b, ~0.2 °C), we can see that more than half of the warming since the 1970s is the result of the timing of the upturn of the 20 and 60 year cycles. Our empirical model further predicts a lack of warming since about 2000, in contrast to IPCC models

In conclusion, we have shown that the effect of natural oscillations is critical for proper assessment of anthropogenic impacts on the climate system. If the rapid increase in temperature in the 1980s and 1990s is due partially to natural cycles, then any model-based estimate of climate sensitivity based on elevated greenhouse gases (and that ignores instrumental data UHI and LULC contamination) will be too high, a point also made by *Scafetta and West* [2008], *Scafetta* [2009, 2010], and *Klyashtorin and Lyubushin* [2003]. The same applies to attribution studies that focus on these decades. Proper consideration of these cycles and of longer natural cycles is also critical for detecting underlying trends. For example, a millenarian or longer period cycles might explain the pattern of the Medieval Warm Period and Little Ice Age [*Loehle and Singer*, 2010].

It is also worth noting that by starting in a cycle trough in 1900 and ending on a cycle peak in 2000, graphical depictions of warming in IPCC and other reports give an exaggerated depiction of the rate of warming, a point noted by *Karlén* [2005] and *Soon et al.* [2004]. By further failing to mention solar effects and warm bias in the instrumental record, it is implied that almost all the warming of about 0.8°C over this period is human caused, when really no more than 0.4°C is likely to be. Reporting such a lower number would probably cause much less alarm than what was recently claimed by *Rockström et al.* [2009]. Moreover, because the 60-year natural cycle will be in its cooling phase for the next 20 years, global temperatures will probably not increase for the next few decades in spite of the important role of human emissions (see Fig. 5a), as predicted by multiple studies [reviewed in *Loehle*, 2009a; *Scafetta*, 2010].

**Acknowledgements**

No outside funding was used to conduct this work.

**Table 1:** Regression coefficients of the harmonic model plus linear trend depicted in Eq. (1) used in Figure 1 (fit 1850-2010) and in Figure 2 (1850-1950), respectively. Note that the harmonic coefficients A, B, $T_1$ and $T_2$ are compatible within the error of measure. Parameter A is the 60-year cycle amplitude and B the 20-year cycle amplitude.

|       | Case 1 | Case 2 |
|-------|--------|--------|
| A     | 0.121 +/- 0.015 ºC | 0.121 +/- 0.016 ºC |
| B     | 0.034 +/- 0.016 ºC | 0.041 +/- 0.016 ºC |
| C     | 0.0042 +/- 0.0002 ºC/year | 0.0016 +/- 0.0004 ºC/year |
| D     | -0.297 +/- 0.013 ºC | -0.317 +/- 0.011 ºC |
| $T_1$ | 1998.62 +/- 1.2 year | 1998.58 +/- 1.3 year |
| $T_2$ | 2000.98 +/- 1.5 year | 1999.65 +/- 1.3 year |

**Figure Legends**

Figure 1.    a) Model for entire period utilizing HadleyCRUT3 global surface temperature April 8, 2010 dataset.  b) Residuals showing clear model mis-specification.

Figure 2.    a) As in Figure 1a, with model fit to pre-1950 data.  b) Residuals. Before about 1950 residuals are stationary around the zero level. After about 1942 there is a clear upward linear trend which may be associated to anthropogenic warming.

Figure 3.    a) Full reconstruction model, which uses Eq. 1 (= natural modulation) fit from 1850 to 1950 plus the additional residual upward linear trend (=anthropogenic modulation) since 1942.  b) Components of the model.

Figure 4.    a) *G. Bulloides* abundance variation record found in the Cariaco Basis sediments in the Caribbean sea since 1650 [Black et al., 1999]. b) tree-ring chronologies from Pinus Flexilis [MacDonald and Case, 2005] as an index to the PDO. Both records show five large quasi 60-year cycles since 1650.

Figure 5.    Forecast models.  Thick line is the global surface temperature. Dashed thick line is the total (natural plus anthropogenic) model reconstruction.  Dashed light line is the continuation of the modeled natural component alone.

Figure 6.    Global surface temperature against the phenomenological solar signature obtained with the model by *Scafetta* [2009] applied to the green and black TSI proxy model curves depicted in Supplemental Information 2.

Figure 1

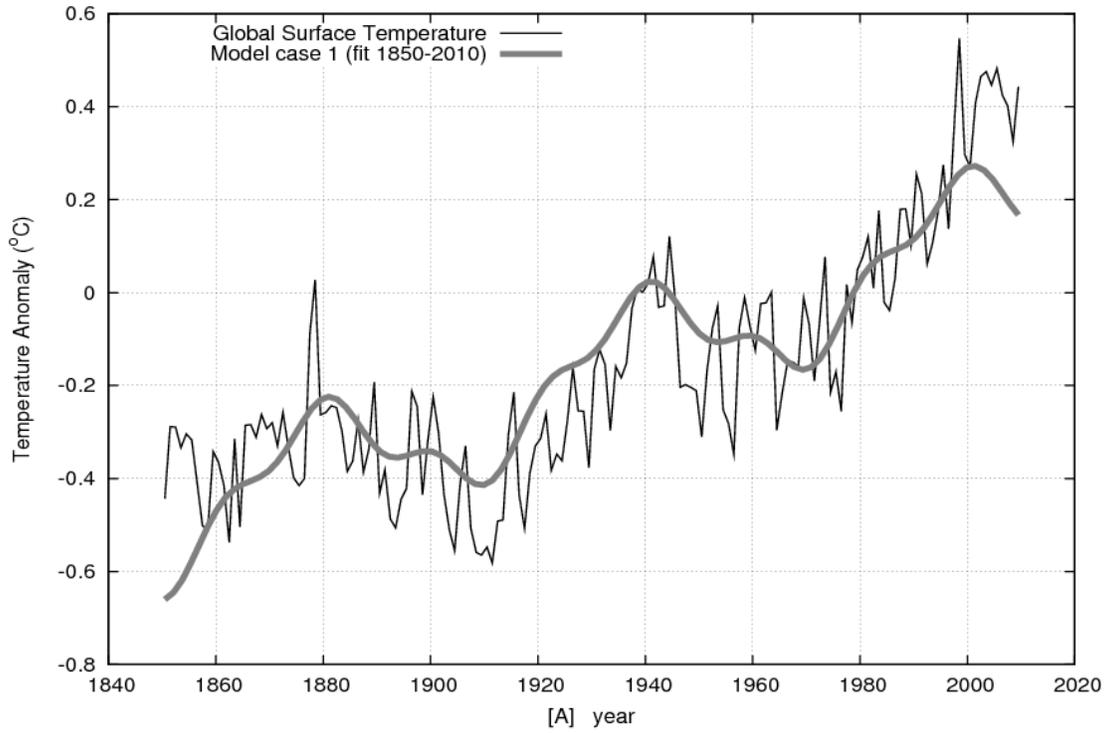

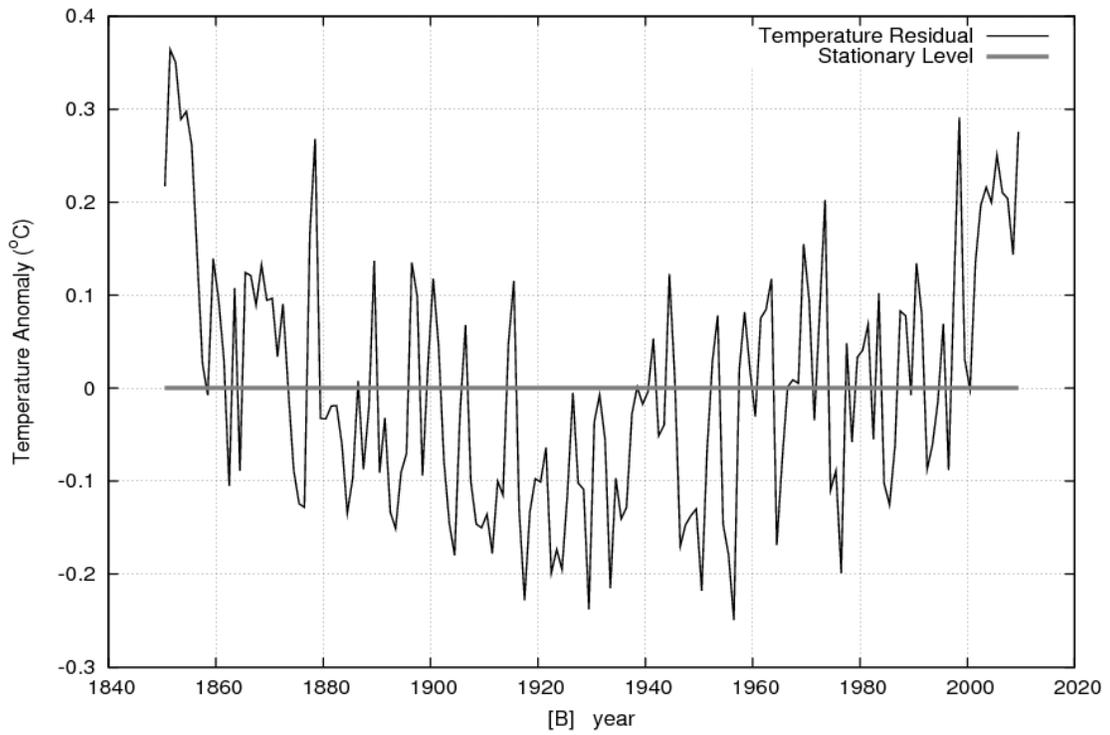

Figure 2.

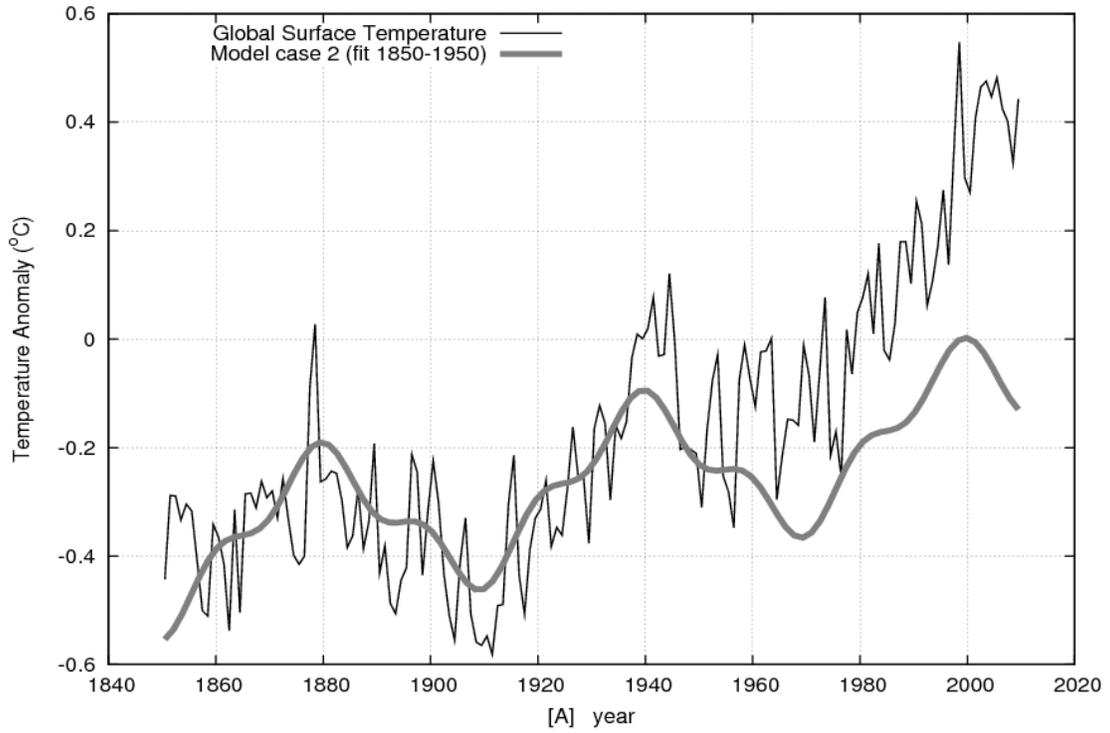

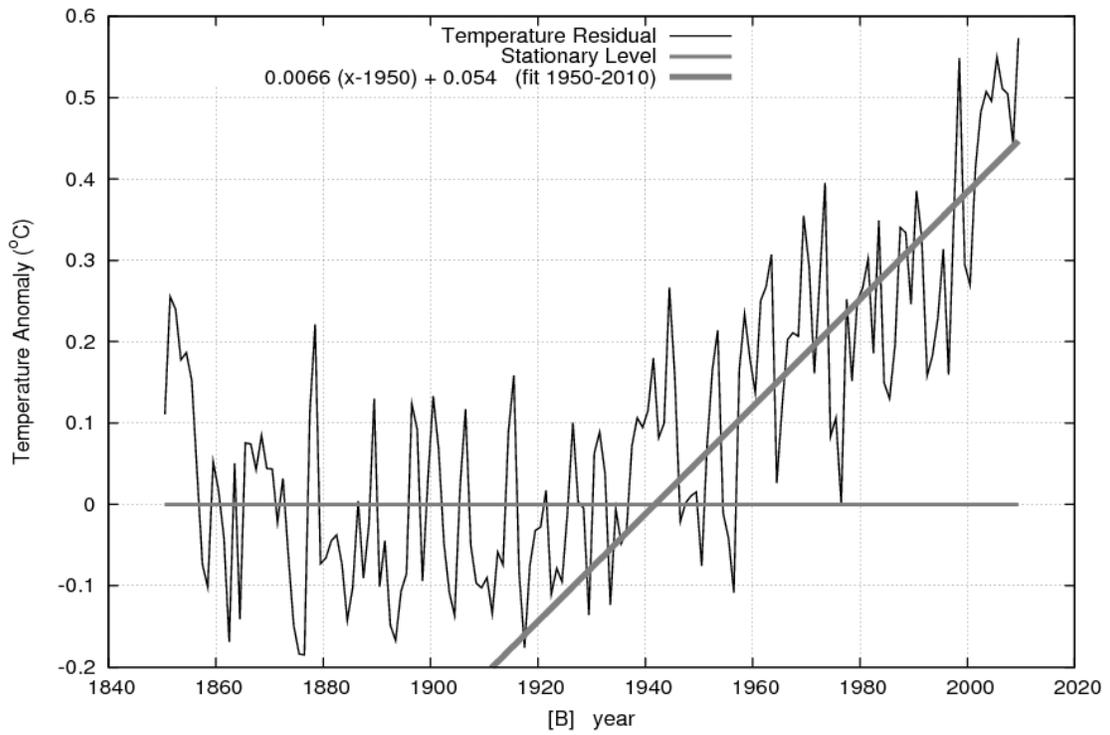

Figure 3.

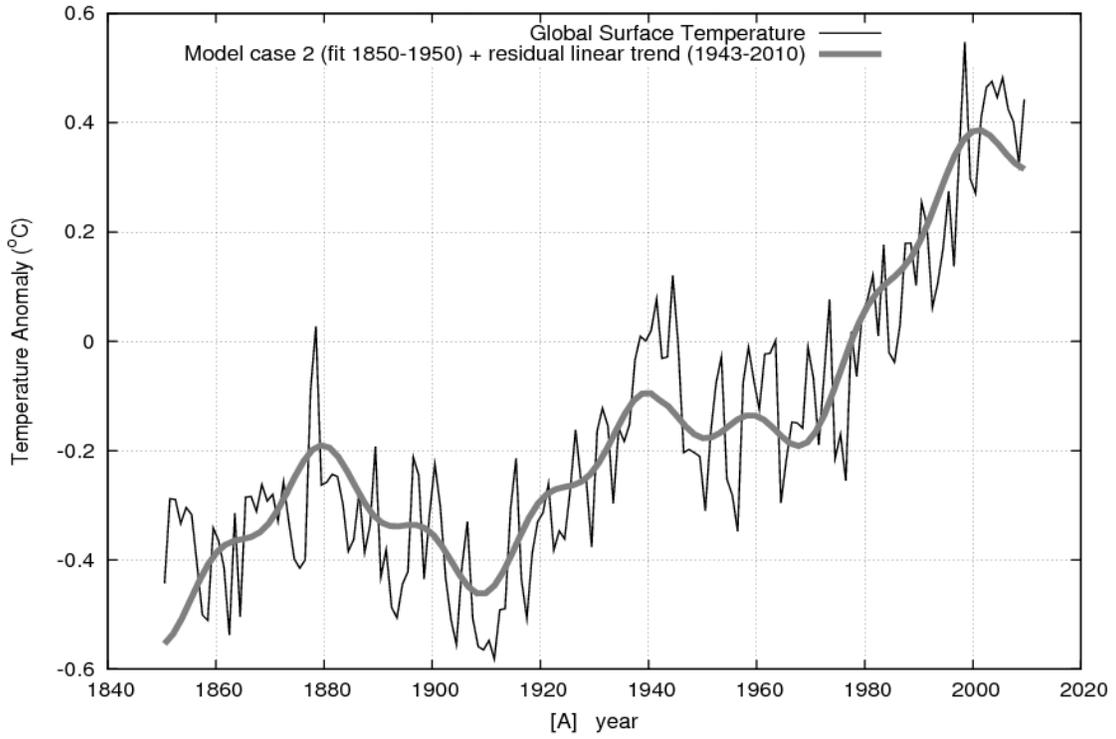

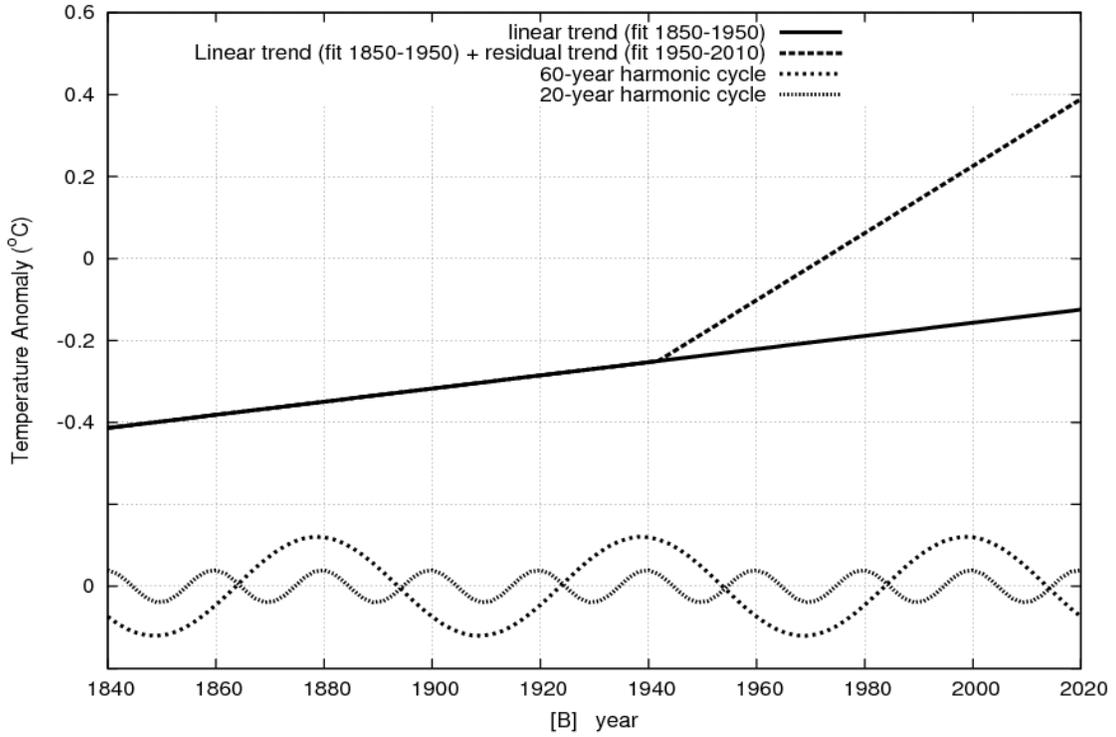

Figure 4.

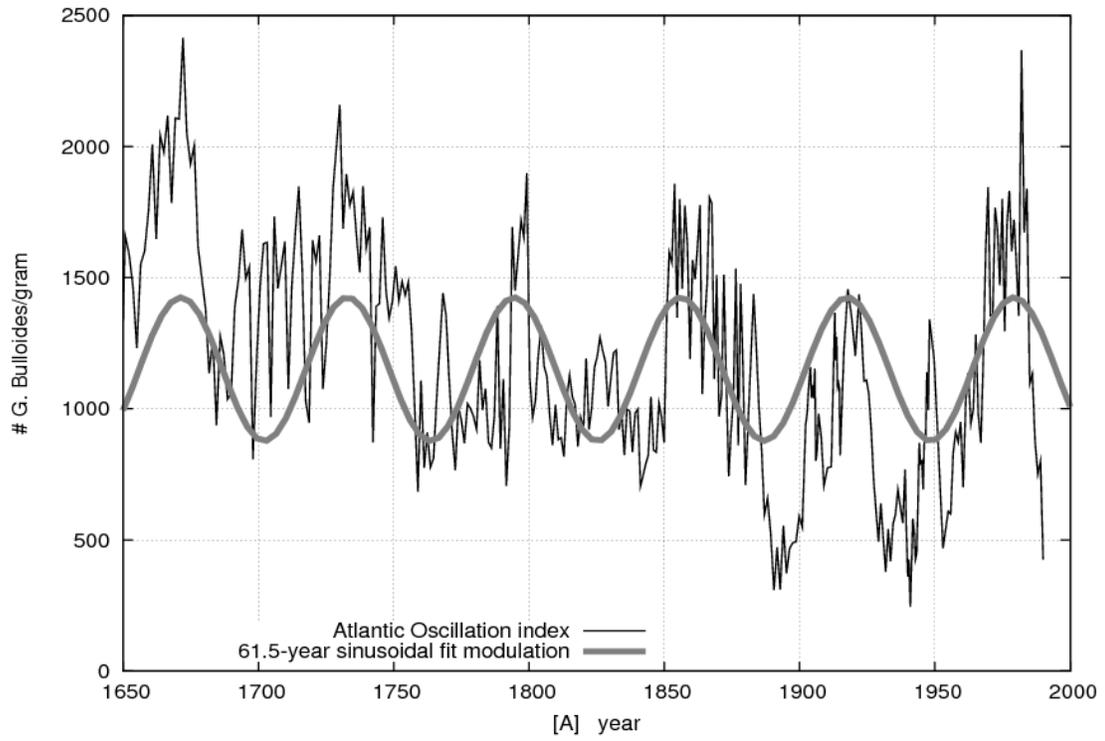

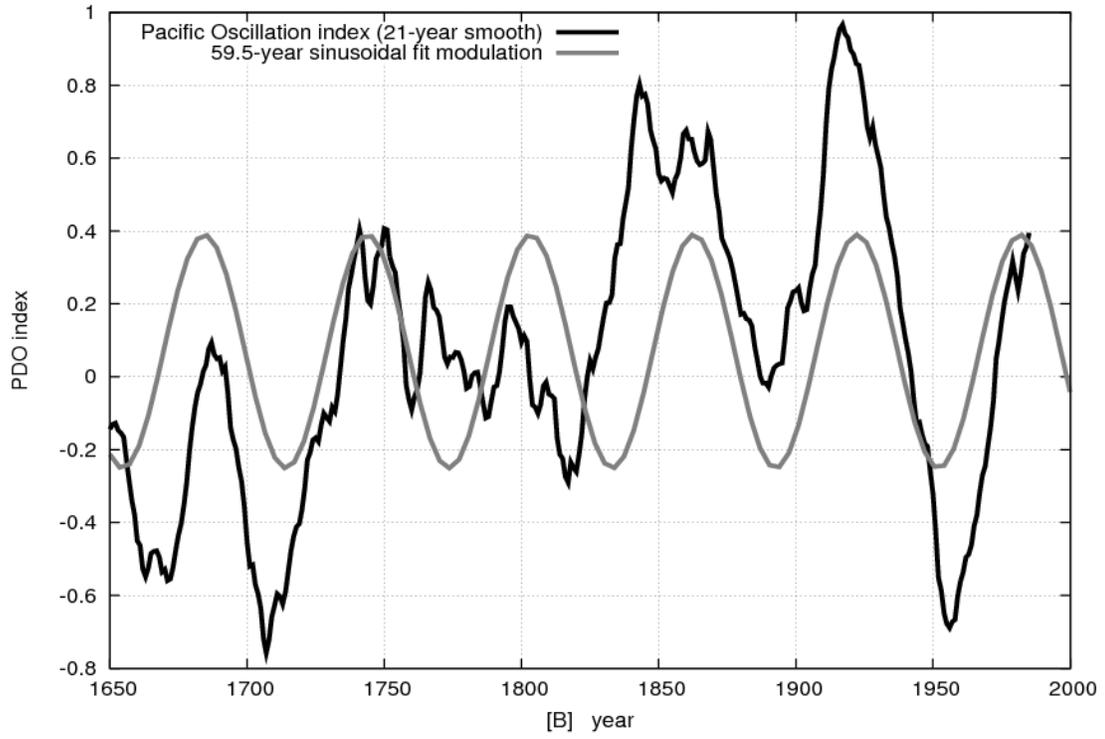

Figure 5

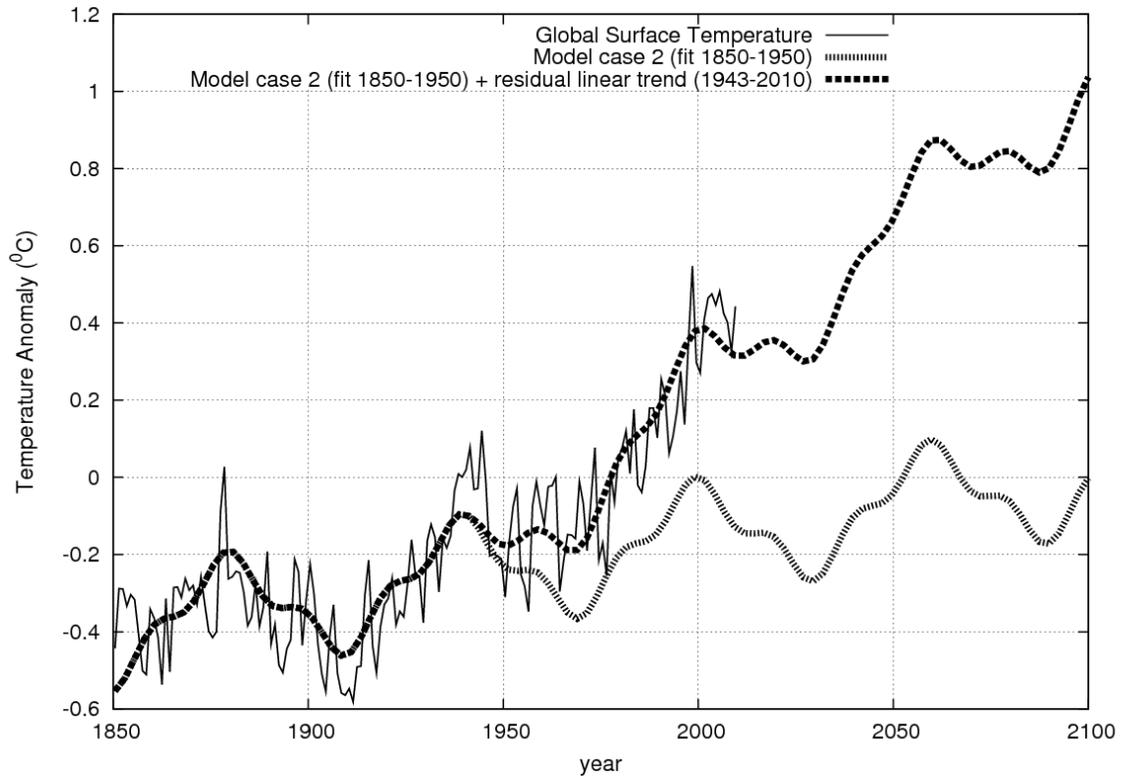

Figure 6

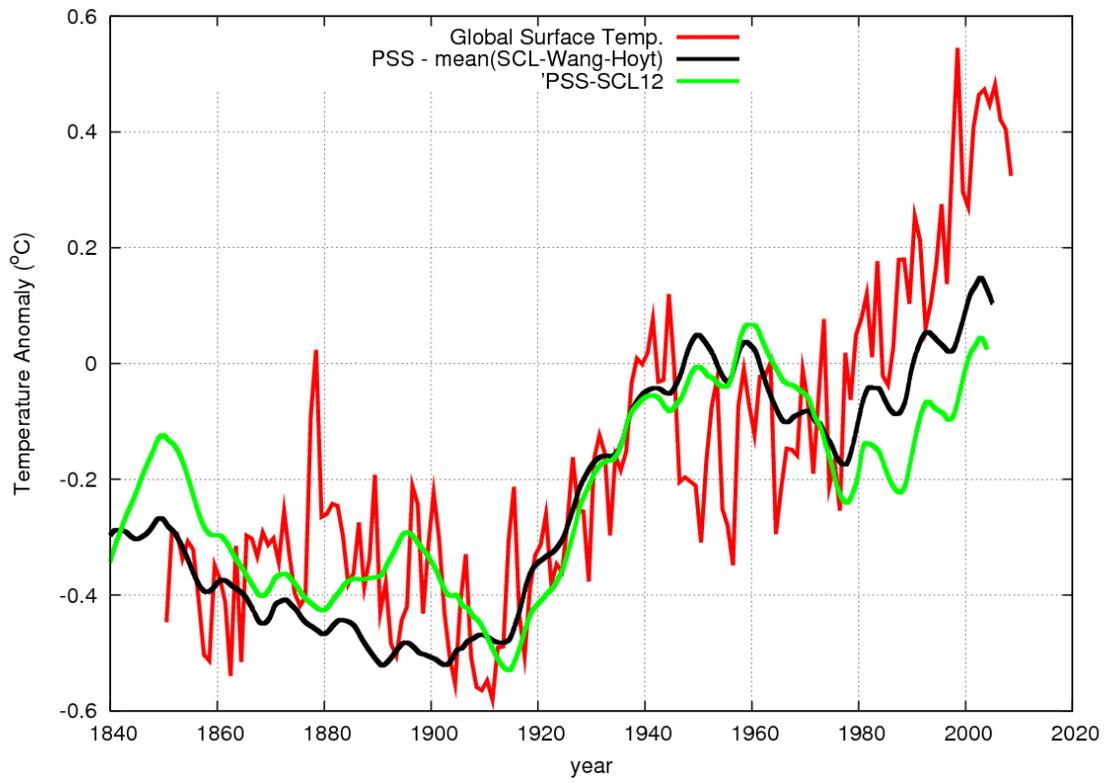

**Climate change attribution using empirical decomposition of climatic data**

∽∽Supplemental Information∽∽

**Nicola Scafetta and Craig Loehle**

**a) A statistical test of the determination of a ~60-year cycle in a short record**

**b) Discussion of solar proxy and satellite composite models**

**c) An empirical climate model with short and long characteristic time responses**

**a) A statistical test of the determination of a ~60-year cycle in a short record**

It can be argued that estimation of periodic signals in climate data requires either spectral methods or very long time series, or both.  For obtaining a very precise estimate of signal properties this may be true, but for estimating periodicity to within plus or minus a few years, this is not true.  To illustrate, we replicate the procedure used in the main manuscript.  We take the model from Fig. 1 as the target.  A sample set is obtained by randomly adding white noise to this model with variance derived from Fig. 1b (sd = 0.13°C).  The new model is estimated for the period 1850 to 1950 and then extrapolated to 2009 (see Fig. Supp. 1).  Repeating this 1000 times, the mean estimated period is 64.11 yr vs. 64.93 for the target, with sd = 2.63 yr.  Thus, clearly the proposed estimation method works sufficiently well even on a 100 yr calibration period.

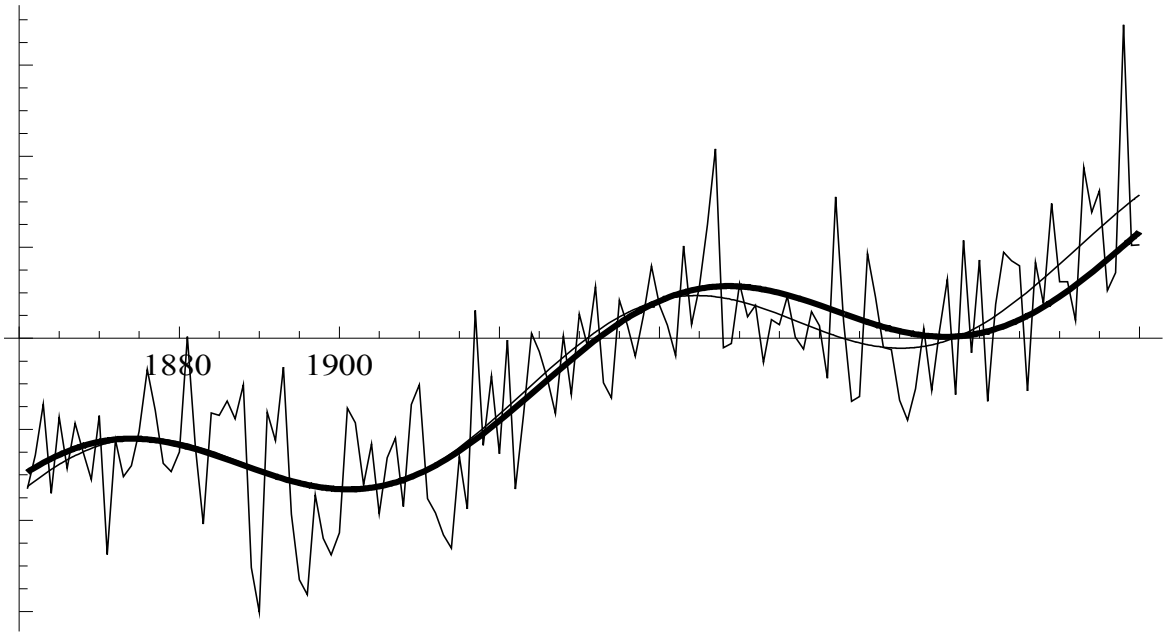

Figure Sup. 1.   Example of model estimation.  Thin line is target cycle.  Jagged line is target with sample white noise added.  Thick line is model fit to randomized data up through 1950 and then extrapolated over the post-1950 period.

**b) Discussion of solar proxy and satellite composite models**

The first direct measurements of TSI were possible in 1978, the year the first satellite experiment measuring TSI started. However, the contiguous 30 year TSI database of satellite observations from late 1978 to the present has to be reconstructed because no satellite record covers the entire period. Two major groups using TSI satellite measurements have proposed alternative composites. The ACRIM composite [*Willson and Mordvinov*, 2003] shows an upward trend during the period 1980-2002, while the PMOD composite [*Frohlich and Lean*, 1998; *Frohlich*, 2006] shows an almost constant trend during the same period. This paper is not the place for a detailed discussion of the relative merits of the PMOD and ACRIM composites; such comparisons have been made elsewhere [e.g., *Scafetta and Willson*, 2009]. Only proxy models can be used for the period prior to 1978. These models adopt ground observables that can reasonably be related to solar activity. Some of the records used are sunspot number, solar rotation, the width of a few spectral lines (10.7 cm radio flux, CaII index, MgII index, HeI index, and a few others), a few magnetic indices, and the radioisotopes $^{14}$C and $^{10}$Be that are also used as proxies for solar activity [*Pap and Fox*, 2004]. All these indexes can give only a very partial representation of the variability of solar activity. In our empirical treatment, we view TSI as just a proxy for all solar components that contribute forcing to the earth system, which include all frequencies of the spectrum (visible light, UV, IR, radio), cosmic ray modulation effects and other possible solar induced effects not well identified yet.

Depending on which proxies one chooses and how one uses them, different TSI proxy models emerge. For example, the TSI proxy models suggested by Lean and colleagues [*Lean et al.*, 1995; *Lean*, 2000; *Wang et al.*, 2005] are very different from each other and differ significantly from the TSI proxy models proposed by other authors [*Hoyt and Schatten*, 1997; *Krivova et al.*, 2007]. In addition, a TSI model based on solar cycle length has been also proposed [*Friis-Christensen and Lassen*, 1991; *Thejll and Lassen*, 2000].

Although these proxy models look vaguely similar to each other (e.g., all of them show a minimum during the Maunder (1650-1715) and Dalton (1790-1830) Minima and a maximum during the last decades), the detailed patterns differ considerably. Figure 2 compares three independent TSI reconstructions: *Wang et al.* [2005] (red); *Hoyt and Schatten* [1997] (blue); and our revision of a TSI proxy model proposed by *Thejll and Lassen* [2000] (green).

The model proposed by *Wang et al.* [2005] (red) shows a slight decrease from 1850 to 1910, an increase from 1910 to 1960, a slight decrease followed by a slight increase from 1960 to 1980, and a constant behavior from 1980 to 2000, which is compatible with PMOD.

The model proposed by *Hoyt and Schatten* [1997] (blue) presents a constant behavior from 1850 to 1880, a decrease from 1880 to 1890, an increase from 1890 to 1940/45, a decrease from 1940/45 to 1970, and an increase from 1970 to 2000, which is compatible with ACRIM. Note that herein we are interested in this pattern structure of increasing and decreasing TSI trends, more than the actual absolute amplitude of the signal, that is an independent issue.

The TSI proxy model (green) is a revision of the proposed model by *Thejll and Lassen* [2000] based on solar cycle length. The physical meaning of this model is that when the solar cycle is longer the sun is quieter because the solar dynamo is running slower, so its irradiance output is lower.

Our revision of the solar cycle length model is made as follows:

(1)    We used the same solar cycle length published in *Thejll and Lassen* [2000] but correct the last cycle length (originally predicted to cover the period from 1996.8 to 2007.3) to the observed period from 1996.8 to 2009.

(2)    We applied a type 1-2 smoothing algorithm to extract the trend; that is, we use the algorithm $S_i = (A_{i-1} + 2A_i)/3$ (see Table 1).  This is different from what was suggested by *Thejll and Lassen*, a centered smoothing of the type 1-2-1 and 1-2-2-2-1.  There are two reasons for our choice:  (a)  the algorithm proposed by *Thejll and Lassen* may not be physical because they assume that the present behavior of the sun may be conditioned by future cycles; and (b) our choice of a smoothing of type 1-2 takes into account that a solar trend may be part of two consecutive 11 year solar cycles that are known to be linked by the 22 year Hale magnetic solar cycle.  Moreover, it is assumed that one 11-year solar cycle feels a memory from the previous cycle as all physical complex systems tend to do. Note that this way of doing the calculation avoids the shortcomings of the original records published by *Friis-Christensen and Lassen* [1991] and updated by *Thejll and Lassen* [2000] due to end point issues for which those papers have been criticized.

(3)    The 1-2 smoothed solar cycle length was transformed into a smooth curve with an annual resolution with an interpolation cspline algorithm.

(4)    The curve was inverted because short cycles would imply higher TSI.  Because the inverted curve presents an increase from 1980 to 2000 that matches the ACRIM record, we calibrate the model to reproduce the ACRIM trend increase from 1980 to 2000.

(5)    The 11-year modulation extracted from *Wang et al.* [2005] was added to the 1-2 smoothed solar cycle length.

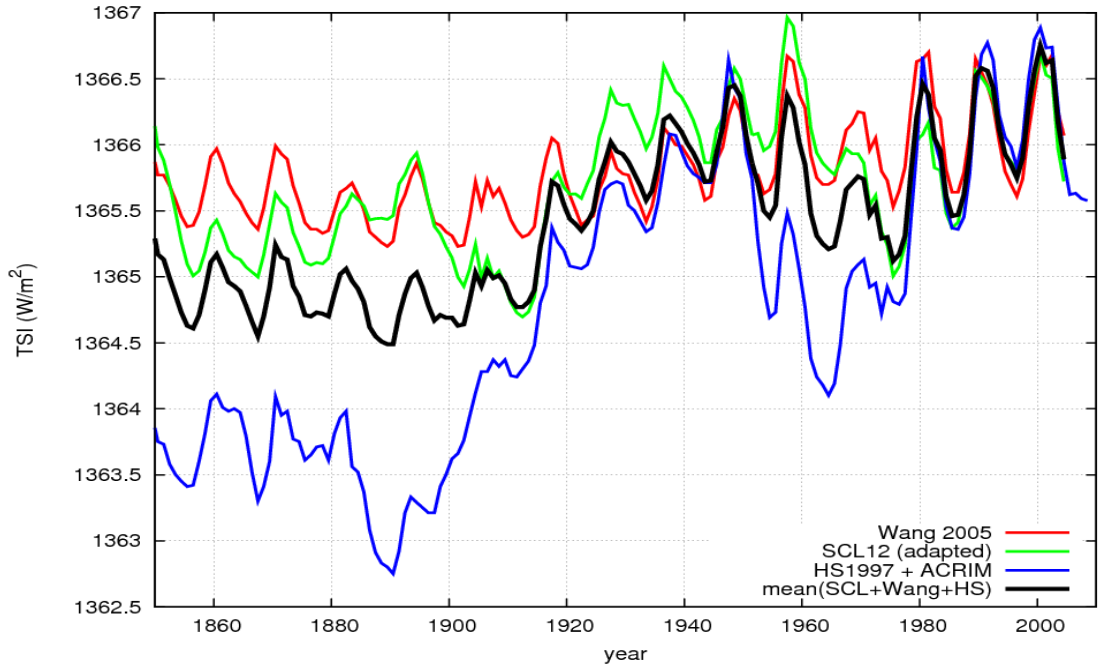

Figure Sup. 2. Three independently constructed TSI proxy models. *Wang et al.* [2005] (red); *Hoyt and Schatten* [1997] (blue) recalibrated at ACRIM level; and a revision of a TSI proxy model proposed by *Thejll and Lassen* [2000] (green) based on a 1-2 smooth solar cycle length. The black curve is the average among the three models.

Table 1.  Corrected smoothed solar cycle data based on *Thejll and Lassen* [2000]. [a] cycle number and year of solar minima at the end of the cycle from *Thejll and Lassen*, 2000; last value corrected to 2009. [b] solar cycle length; [c] solar cycle trend according to 1-2 role smooth

| Solar Min Year[a] | SCL[b] | 1-2 SCL[c] |
|---|---|---|
| 00 – 1755.2 | 11.2 | |
| 01 – 1766.5 | 11.3 | 11.27 |
| 02 – 1775.5 | 9 | 9.77 |
| 03 – 1784.7 | 9.2 | 9.13 |
| 04 – 1798.3 | 13.6 | 12.13 |
| 05 – 1810.6 | 12.3 | 12.73 |
| 06 – 1823.3 | 12.7 | 12.57 |
| 07 – 1833.9 | 10.6 | 11.3 |
| 08 – 1843.5 | 9.6 | 9.93 |
| 09 – 1856.0 | 12.5 | 11.53 |
| 10 – 1867.2 | 11.2 | 11.63 |
| 11 – 1878.9 | 11.7 | 11.53 |
| 12 – 1889.6 | 10.7 | 11.03 |
| 13 – 1901.7 | 12.1 | 11.63 |
| 14 – 1913.6 | 11.9 | 11.97 |
| 15 – 1923.6 | 10 | 10.63 |
| 16 – 1933.8 | 10.2 | 10.13 |
| 17 – 1944.2 | 10.4 | 10.33 |
| 18 – 1954.2 | 10 | 10.13 |
| 19 – 1964.9 | 10.7 | 10.47 |
| 20 – 1976.5 | 11.6 | 11.3 |
| 21 – 1986.8 | 10.3 | 10.73 |
| 22 – 1996.8 | 10 | 10.1 |
| 23 – 2009.0 | 12.2 | 11.47 |

**c) An empirical climate model with short and long characteristic time responses**

Herein we briefly summarize, for the convenience of the reader, the empirical model used by *Scafetta* [2009] for reconstructing the solar signature on climate, which has been used to produce Fig. 6 in the main report. The model assumes that the climate system processes an external input forcing exactly like a traditional climate energy balance model. That is, it is assumed that the thermodynamical properties of climate are characterized by a given sensitivity to the external forcing and by a given heat capacity. The latter determines the time response of the system. The difference between the empirical model and the more traditional energy balance models is that both the climate sensitivity to an external forcing and the time responses of the system are empirically measured on the climate data themselves instead of being theoretically *a priori* deduced.

By using auto-correlation properties of the temperature, *Scafetta* [2008] determined that the climate system is characterized by at least two major characteristic time constants which are about: $\tau_1 = 0.4$ year and $\tau_2 = 12$ year. These two time constants are physically reasonable because they agree well with the idea that the climate system can be approximately assumed to be made of two coupled subsystems such as, for example, the air-land and the ocean. The former subsystem (air+land) is characterized by a relatively small heat capacity; therefore, it responds quickly to an external forcing. The latter subsystem (ocean) is characterized by a relatively large heat capacity; therefore, it responds slowly to an external forcing.

Thus, climate can be approximately considered a superposition of two subsystems each with its own characteristic time response and sensitivity. The two climate sensitivities were deduced in *Scafetta* [2009] by taking into account that the overall climate response to the 11 year solar cycle induces a cycle with a peak-to-trough signature on the global surface temperature of about 0.1°C, as found by several authors [*Scafetta*, 2009; *IPCC*, 2007 p 674].

The measured climate sensitivities to solar variation were $k_{1s} = 0.053$ K/Wm$^{-2}$ and $k_{2s} = 0.41$ K/Wm$^{-2}$ for the two subsystems, respectively. Note that these sensitivities are not equivalent to the climate sensitivities to radiative forcing as used by traditional climate models such as those used by the *IPCC* [2007], they are larger. In fact, in the present case, the total solar irradiance is used just as a proxy for the overall climate sensitivity to the overall solar forcing that is not just TSI alone. For example, there may be an additional strong contribution from solar modulated cosmic ray flux that can modulate cloud formation.

By using the above values, the empirical solar signature on climate, which is indicated by $\Delta T(t)$, associated to a given TSI record, which is indicated by $\Delta I(t)$, can be calculated from the following system of differential equations:

$$
\begin{cases}
\Delta T = \Delta T_{1S} + \Delta T_{2S} \\
\dfrac{d\Delta T_{1S}(t)}{dt} = \dfrac{k_{1S}\Delta I(t) - \Delta T_{1S}(t)}{\tau_1} \\
\dfrac{d\Delta T_{2S}(t)}{dt} = \dfrac{k_{2S}\Delta I(t) - \Delta T_{2S}(t)}{\tau_2}
\end{cases}
\tag{3}
$$

These equations were used to derive the curves depicted in Fig. 6 in the main report. More details are found in *Scafetta* [2009].